\documentclass{nature}
\usepackage{latexsym}
\usepackage{amsthm}
\usepackage{amssymb}
\usepackage{amsmath}
\usepackage[all]{xy}
\usepackage{graphicx}
\usepackage{subfigure}
\usepackage{dcolumn}
\usepackage{bm}
\usepackage{bbm}
\usepackage[usenames,dvipsnames]{xcolor}
\usepackage{amsfonts}
\usepackage{cases}
\usepackage[a4paper,colorlinks=true,pagebackref=true,
linkcolor=blue,citecolor=blue,
pdfauthor={ },
pdftitle={ },
pdfsubject={ },
pdfkeywords={ }]{hyperref}
\newcommand{\ket}[1]{|#1\rangle}
\newcommand{\bra}[1]{\langle#1|}

\begin{document}
\global\long\def\s{\sigma}
 \global\long\def\k{\kappa}
 \global\long\def\ph{\hat{n}}
 \global\long\def\aa{\hat{a}}
 \global\long\def\bra#1{\left\langle #1\right|}
 \global\long\def\ket#1{\left|#1\right\rangle }

\title{Direct measurement of topological numbers with spins in diamond}


\author
{Fei Kong,$^{1\dag}$
Chenyong Ju,$^{1,2\dag}$
Ying Liu,$^{1\dag}$
Chao Lei,$^{1,2}$
Mengqi Wang,$^{1}$
Xi Kong,$^{1,2}$
Pengfei Wang,$^{1,2}$
Pu Huang,$^{1,2}$
Zhaokai Li,$^{1,2}$
Fazhan Shi,$^{1,2}$
Liang Jiang,$^{3\ast}$
Jiangfeng Du$^{1,2}$\thanks{e-mail: liang.jiang@yale.edu; djf@ustc.edu.cn $^{\dag}$These authors contributed equally to this work.}\\}

\maketitle

\begin{affiliations}
\item
National Laboratory for Physical Sciences at the Microscale and Department of Modern Physics, University of Science and Technology of China, Hefei, 230026, China
\item
Synergetic Innovation Center of Quantum Information and Quantum Physics, University of Science and Technology of China, Hefei, 230026, China
\item
Department of Applied Physics, Yale University, New Haven, Connecticut 06511, USA

\end{affiliations}

\begin{abstract}
Topological numbers can characterize the transition between different topological phases, which
are not described by Landau's paradigm of symmetry breaking. Since
the discovery of quantum Hall effect, more topological phases
have been theoretically predicted and experimentally
verified. However, it is still an experimental challenge to directly
measure the topological number of various predicted topological phases.
In this paper, we demonstrate quantum simulation of topological phase transition of a
quantum wire (QW) using a single nitrogen-vacancy (NV) center in diamond. Deploying quantum algorithm of finding
eigenvalues, we can reliably extract both the dispersion relations and topological numbers.
\end{abstract}
\maketitle

Topological numbers were first introduced by Dirac to justify the
quantization of electric charge \cite{Dirac1931}, and later developed
into a theory of magnetic monopoles as topological defects of a gauge
field \cite{Dirac1948}. An amazing fact is that fundamental quantized
entities may be deduced from a continuum theory \cite{thouless1998}. Later on, topological
numbers were used to characterize the quantum Hall effect \cite{Klitzing1980,Stormer1999}
in terms of transition between topological phases \cite{wen1990}. Since the
topological number in quantum Hall systems is directly proportional to the resistance in
transport experiment, its robustness against local perturbations enables a practical standard for electrical resistance \cite{Klitzing1980}. In the past few years, more topological
materials have been discovered, including topological insulators\cite{Qi2011,Hasan2010},
topological superconductor\cite{alicea2012new,Beenakker2015},
and etc.

Developing robust techniques to probe topological numbers becomes
an active research topic of both fundamental and practical importance.
Recently, a generalized method of extracting topological number by integrating
dynamic responses has been proposed \cite{Gritsev12}. Guided by this theoretical
proposal, experiments have successfully measured the topological Chern number
of different topological phases using superconducting circuits \cite{Schroer2014,roushan2014}.
However, their measurement of Chern number requires integration over
continuous parameter space, which may not give an exactly discretized topological number. Different from the above integration approach, here we take the simulation approach \cite{feynman1982,lloyd1996,Georgescu2014} and use a single NV center in natural diamond at room temperature \cite{Gruber97,Maurer08062012} to simulate a topological system.
Moreover, we deploy quantum algorithm of finding eigenvalues to map
out the dispersion relations \cite{Ju2014,EigenAlgorithm}
and directly extract the topological number, which enables direct observation of the simulated topological phase transition.

We consider the topological phase transition associated with a semiconductor
quantum wire with spin-orbital interaction, coupled to s-wave superconductor
and magnetic field \cite{Oreg10,Lutchyn10,alicea2012new,Jiang2013_Theory}.
At the boundary between different topological phases of the quantum
wire, Majorana bound states can be created as a promising candidate
for topological quantum information processing \cite{Alicea11}. The
Hamiltonian of this system can be described using Nambu spinor basis
${\psi^{T}}=\left(\psi_{\uparrow},\psi_{\downarrow},\psi_{\downarrow}^{\dag},-\psi_{\uparrow}^{\dag}\right)$:
\begin{equation}
{H_{\text{QW}}}=p{\sigma_{z}}{\tau_{z}}+\left({{p^{2}}-\mu}\right){\tau_{z}}+\Delta{\tau_{x}}+{B_{x}}{\sigma_{x}},\label{QW_Hamiltonian}
\end{equation}
with the momentum $p$, chemical potential $\mu$,
pairing amplitude $\Delta$, Zeeman energy $B_{x}$, and Pauli matrices
$\sigma_{a}$ and $\tau_{a}$ acting in the spin and particle-hole
sectors respectively. Without loss of generality, we may assume negative
$\mu$, non-negative $B_{x}$ and $\Delta$.

The system described by equation~(\ref{QW_Hamiltonian}) has two different
topological phases determined by the relative strength of $\{B_{x},\mu,\Delta\}$:
(i) the trivial superconductivity phase (denoted by SC phase) when
${B_{x}}<\sqrt{{\Delta^{2}}+{\mu^{2}}}$, and (ii) the topological
superconductivity phase (denoted as TP phase) when ${B_{x}}>\sqrt{{\Delta^{2}}+{\mu^{2}}}$.
The phase diagram and dispersion relations of different phases are
illustrated in Fig.~\ref{Phase_diagram}a. There are four energy bands
for this system, consisting of two particle bands and two hole bands.
We may label the energy bands as $1,2,3,4$ from bottom to up as in
Fig.~\ref{Phase_diagram}a. The gap between the 2nd and 3rd bands
will disappear during the phase transition.
To illustrate the distinct topological nature
associated with the SC and TP phases, we may consider the transformed
Hamiltonian $\tilde{H}_{QW}=U_{p}H_{QW}U_{p}^{\dagger}$ with unitary
transformation $U_{p}=\exp[i\theta_{p}\cdot\tau_{y}]$ and $\theta=\frac{1}{2}\arctan\frac{\Delta}{p^{2}-\mu}$,
which preserve eigenenergies and all energy bands. Then, we plot the normalized Bloch vector of the reduced eigenstate of the 3rd
band in the Bloch sphere associated with the particle-hole sector
(i.e., taking a partial-trace over the spin sector for the 3rd eigenstate).
As shown in Fig.~\ref{Phase_diagram}b, when the momentum $p$ changes
from 0 to $\infty$, the trajectory of the state vector for the SC phase forms
a closed loop, while for the TP phase it remains an open trajectory
starting from one pole and ending at the opposite pole. Here we treat
$p\rightarrow-\infty$ and $p\rightarrow+\infty$ as identical to one another, as $H_{\text{QW}}\rightarrow p^{2}\tau_{z}$
for the both cases. This implies an inversion of the particle and
hole bands for the TP phase (but not for the SC phase) with increasing
momentum from $p=0$ to $p\rightarrow\infty$. The two types of topologically
different trajectories imply the existence of distinct
topological phases for the system. Mathematically, a topological number
can be introduced to distinguish these two phases:
\begin{equation}
\nu\equiv{\mathop{\rm sgn}}[\mathrm{Pf}(H_{\text{QW}}(p=0))]\cdot{\mathop{\rm sgn}}[\mathrm{Pf}(H_{\text{QW}}(p\rightarrow\infty))],\label{Topo_invariant}
\end{equation}
where $\mathrm{Pf}$ is the Pfaffian\cite{Kitaev2001}
with $\mathrm{Pf}(H_{\text{QW}}(p))=\mathrm{Pf}(\tilde{H}_{\text{QW}}(p))$
for $U_{p}^{\dagger}=U_{p}^{T}$.
Specifically, $\mathrm{Pf}(H_{\text{QW}}(p=0))=-B_{x}^{2}+\mu^{2}+\delta^{2}$ and
$\mathrm{Pf}(H_{\text{QW}}(p\rightarrow\infty))=p^{4}$, the value of $\nu$
is determined by the sign of the quantity $-B_{x}+\sqrt{{\mu^{2}}+{\Delta^{2}}}$,
which is one of eigenenergies of $H_{\text{QW}}(p=0)$. The corresponding
eigenstate can be represented by $\left|\Phi\right\rangle =\left|\Phi_{\sigma}\right\rangle \bigotimes\left|\Phi_{\tau}\right\rangle $,
where $\left|\Phi_{\sigma}\right\rangle $ and $\left|\Phi_{\tau}\right\rangle $
are the eigenstates of ${B_{x}}{\sigma_{x}}$ and $-\mu{\tau_{z}}+\Delta{\tau_{x}}$
with eigenenergies $-B_{x}$ and $\sqrt{{\mu^{2}}+{\Delta^{2}}}$
respectively. It is direct to deduce $\left|\Phi_{\sigma}\right\rangle =\left|\leftarrow\right\rangle =(\left|\uparrow\right\rangle -\left|\downarrow\right\rangle )/\sqrt{2}$
($\left|\uparrow\right\rangle $ and $\left|\downarrow\right\rangle $
means spin up and down) and $\left|\Phi_{\tau}\right\rangle =\alpha\left|p\right\rangle +\beta\left|h\right\rangle $
($\left|p\right\rangle $ and $\left|h\right\rangle $ means particle
and hole) which is dominated by $\left|p\right\rangle $ (i.e. $|\alpha|^{2}>|\beta|^{2}$).
Therefore, we can apply quantum algorithm of finding eigenvalues \cite{EigenAlgorithm}
for the state $\left|\Phi\right\rangle $ to directly obtain the eigenenergy
$-B_{x}+\sqrt{{\mu^{2}}+{\Delta^{2}}}$, the sign of which is exactly
the topological number $\nu$.

The QW Hamiltonian is simulated by a highly controllable two-qubit solid-state system, which is a color defect named NV center in diamond consisting of a substitutional nitrogen
atom and a adjacent vacancy, as shown in Fig.~\ref{Protocol}a. The electrons around the defect form an effective electron spin with a spin triplet ground state ($S=1$) and couple with the nearby $^{14}$N nuclear spin.
With an external magnetic field $B_{0}$ along the
N-V axis, the Hamiltonian of the NV system is ($\hbar=1$) \cite{Loubser1978RPP}:
\begin{equation}
H_{\text{NV}}^{0}=-{\gamma_{\text{e}}}{B_{0}}{S_{z}}-{\gamma_{\text{n}}}{B_{0}}{I_{z}}+DS_{z}^{2}+QI_{z}^{2}+A{S_{z}}{I_{z}},\label{NV0_Hamiltonian}
\end{equation}
where $S_{z}$ and $I_{z}$ are the spin operators of the electron
spin (spin-1) and the $^{14}$N nuclear spin (spin-1), respectively.
The electron and nuclear spins have gyromagnetic ratios $\gamma_{\text{e}}/2\pi=-28.03$
GHz/T and $\gamma_{\text{n}}/2\pi=3.077$ MHz/T, respectively. $D/2\pi=2.87$
GHz is the axial zero-field splitting parameter for the electron spin,
$Q/2\pi=-4.945$ MHz is the quadrupole splitting of the $^{14}$N
nuclear spin, and $A/2\pi=-2.16$ MHz is the hyperfine coupling constant.
There are nine energy levels, $\left|1\right\rangle ,\cdots,\left|9\right\rangle $,
as labeled in Fig.~\ref{Protocol}b.
The simulation is performed in the subspace spanned by $\{\left|4\right\rangle ,\left|5\right\rangle ,\left|7\right\rangle ,\left|8\right\rangle \}$,
associated with the electron spin states $\{m_{\text{e}}=0,-1\}$ (encoding the
pseudospin $\sigma$) and the nuclear spin states $\{m_{\text{n}}=0,1\}$
(encoding the pseudospin $\tau$). The NV spins are radiated by two microwave (MW) pulses and
two radio-frequency (RF) pulses simultaneously, which selectively drive the two electron-spin transitions
and the two nuclear-spin transitions respectively (Fig.~\ref{Protocol}b).
While the frequencies of the pulses are all slightly detuned from resonance.
The frequency detunings for the two MW
(RF) pulses are set to be the same value $\delta_{\text{MW}}$ ($\delta_{\text{RF}}$).
In the rotating frame, the Hamiltonian can be written as
(see SI for the detail)
\begin{equation}
\begin{split}H_{\text{NV}}^{\text{rot}}= & \frac{\Omega_{\text{MW1}}-\Omega_{\text{MW2}}}{4}\sigma_{x}\tau_{z}-\frac{1}{2}\delta_{\text{RF}}\tau_{z}\\
 & +\frac{\Omega_{\text{MW1}}+\Omega_{\text{MW2}}}{4}\sigma_{x}+\frac{\Omega_{\text{RF}}}{2}\tau_{x}-\frac{1}{2}\delta_{\text{MW}}\sigma_{z},
\end{split}
\label{NVrot_Hamiltonian}
\end{equation}
where $\Omega_{\text{MW1,2}}$ are the Rabi frequencies of the two
electron spin transitions, $\Omega_{\text{RF}}$ is the nuclear spin
Rabi frequency that is set to the same value for the two nuclear spin
transitions. By choosing $\Omega_{\text{MW1}}=-\Omega_{\text{MW2}}=\Omega_{\text{MW}}$, the parameters for the QW system and the NV spins can
be identified as the following: $p\sim\Omega_{\text{MW}}/2$,
$p^{2}-\mu\sim-\delta_{\text{RF}}/2$,
$\Delta\sim\Omega_{\text{RF}/2}$, and $B_{x}\sim-\delta_{\text{MW}}/2$
\footnote{when performing the experiment, the numerical values of the left side are reduced by a factor of 11 to be coincide with the typical values of NV parameters}.
Hence, $H_{\text{QW}}$ can be exactly reproduced
up to a Hadamard gate on the electron spin transforming the
spin operators $\sigma_{x}\leftrightarrow\sigma_{z}$ in $H_{\text{NV}}^{\text{rot}}$.
The Hadamard gate does not change the eigenvalues and can be fully compensated by modifying the basis states in the experiment.
As shown in Fig.~\ref{Protocol}c, the four states of the NV system can be mapped to QW system one-to-one.

To obtain the energy-dispersion relations of QW and calculate the
topological number, we deploy a quantum algorithm of finding eigenvalues \cite{EigenAlgorithm} to measure the eigenvalues of QW.
The initial state of the NV spins is prepared to $\left(\left|6\right\rangle +\left|\Psi\right\rangle \right)/\sqrt{2}$,
where $\left|6\right\rangle $ is used as a reference state and $\left|\Psi\right\rangle = \left|4\right\rangle,\left|5\right\rangle,\left|7\right\rangle, \text{or}\left|8\right\rangle$.
In general $\left|\Psi\right\rangle $
can be expanded by the QW eigenstates $\left|\Psi\right\rangle =\sum_{j=1}^{4}c_{\psi,j}\left|\phi_{j}\right\rangle $
($H_{\text{QW}}\left|\phi_{j}\right\rangle =E_{j}\left|\phi_{j}\right\rangle $).
By applying the simulating pulses for an adjustable period $m\tau$
($m\in\mathbb{N}$), $\left|\Psi\right\rangle $ evolves under the
effective QW Hamiltonian and accumulates phases $\propto E_{j}m\tau$
with the state becoming $\left(\left|6\right\rangle +\sum_{j=1}^{4}c_{\psi,j}e^{-i2\pi E_{j}m\tau}\left|\phi_{j}\right\rangle \right)/\sqrt{2}$.
It can be transformed back into the representation of NV spin states
($\left|\phi_{j}\right\rangle =\sum_{l=4,5,7,8}c_{l,j}^{*}\left|l\right\rangle $)
and can be written as $\left(\left|6\right\rangle +\sum_{l=4,5,7,8}a_{l,m}\left|l\right\rangle \right)/\sqrt{2}$,
with coefficients $a_{l,m}=\sum_{j=1}^{4}c_{\psi,j}c_{l,j}^{*}e^{-i2\pi E_{j}m\tau}$
which is a function of the QW eigenvalues. The coefficient of $\left|\Psi\right\rangle $,
i.e. $a_{\psi,m}$, can be measured in the experiment. Therefore,
the energy spectrum of the QW Hamiltonian can be obtained by Fourier
transforming of the time-domain signals $\{a_{\psi,m}\}$.
There will be at most four peaks in the energy spectrum with their heights $\varpropto |c_{\psi,j}|^2$. Since the 1st and 4th energy bonds are trivial, we only care about the 2nd and 3rd energy bonds. As $|c_{5(7),2}|^2+|c_{5(7),3}|^2 \gg |c_{4(8),2}|^2+|c_{4(8),3}|^2$ in the case of low momentum $|p| \ll \infty$ (see SI), $\left|\Psi\right\rangle $ can be $\left|5\right\rangle $ or $\left|7\right\rangle $ in the experiment. However, the detection of $\{a_{4,m}\}$ is easier than that of $\{a_{7,m}\}$. By reversing the sign of $\delta_{\text{MW}}$ and $\sigma_{z}$ simultaneously in equation~(\ref{NVrot_Hamiltonian}), one can see that the Hamiltonian remains unchange. It means $\left|\Psi\right\rangle $ can choose $\left|4\right\rangle $ instead of $\left|7\right\rangle $ by using $\delta^{'}_{\text{MW}} = -\delta_{\text{MW}}$.

The experimental realization was preformed on a home-build set-up
which has been described early \cite{Shi2014natphy}. The external
statistic magnetic field was adjusted around 50 mT in order to polarize
the $^{14}$N nuclear spin using dynamic polarization technology \cite{Jacques2009PRL}.
The experimental process is shown in Fig.~\ref{Experiment sequence}a.
At first, the NV system was prepared to $\left|4\right\rangle$ by a 4 $\mu$s laser pulse,
then transformed to the superposition state $(\left|6\right\rangle +\left|\Psi\right\rangle )/\sqrt{2}$
during the initialization process. ($\left|\Psi\right\rangle =\left|5\right\rangle $
by the second row RF pulses and $\left|\Psi\right\rangle =\left|4\right\rangle $
by the third row RF pulses shown in the brackets). After that, the
two RF pulses and the two MW pulses described above were applied simultaneously
with time length $m\tau$ in order
to simulate the QW Hamiltonian. Finally, the state was rotated back to $\left|4\right\rangle$
with phase shift $\theta$ and the photoluminescence was detected. As shown in Fig.\ref{Experiment sequence}b, increasing the $\theta$ would lead
to oscillating photoluminescence.
$a_{\psi,m}$ could be obtained from the oscillation amplitude and phase
(see METHOD for the detail). With different pulse length $m\tau$,
we observed the time-domain evolution of $a_{\psi,m}$ (see
Fig.~\ref{Experiment sequence}c). The eigenvalue of the simulated
Hamiltonian can be acquired by the Fourier transform of this time-domain
signal (Fig.~\ref{Experiment sequence}d).

Fig.~\ref{Result}a shows the energy dispersion relations obtained
in experiment for the two SC points ($\mu=-1.6,-1.44$), two TP points
($\mu=-1.14,-0.98$), and the critical point ($\mu=-1.29$), given
$\Delta=0.165$ and $B_x = 1.3$. The experimental results agree well
with the theoretical expectations except for the TP points. The small energy gap
in TP phase disappears due to the fluctuating magnetic field from the surrounding $^{13}\text{C}$ spin bath,
which induces phase errors on the NV electron spin. The phase errors will cause not only peak broadening but also
peak shifting on the energy spectrum.
In addition, the pulses applied
are not perfectly selective pulse, the crosstalk between these pulses
will also cause slightly peak shift on the energy spectrum. The red
lines in Fig.~\ref{Result}a give the numerical calculated energy
dispersion including these imperfections which nicely coincide with
the experimental results (detailed analysis in SI). Further numerical
simulation suggests that the energy gap can be observed if a NV
sample with longer electron spin coherence time is adopted \cite{ultralong2009}.

Even though the small energy gap of the TP phase is difficult to resolve
at the current experimental condition, the topological number $\nu$
characterizing different topological phases can still be unambiguously
extracted.
As mentioned earlier, $\nu$ can be directly determined by the sign of the eigenenergy of $\left|\Phi\right\rangle $. Since $\left|\Phi\right\rangle $ is dominated by $\left|p,\leftarrow\right\rangle $ which is corresponding to $\left|7\right\rangle $, the eigenenergy of $\left|\Phi\right\rangle $ can be reliably obtained from $\{a_{4,m}\}$.
The sign of the eigenenergy can be calculated as
\begin{equation}
\begin{split}\overline{sgn(E)}=\int_{-\infty}^{+\infty}sgn(E)p(E)dE=\int_{-\infty}^{+\infty}\frac{sgn(E)}{\sqrt{2\pi}\sigma}e^{-\frac{(E-E_{c})^{2}}{2\sigma^{2}}}dE,\end{split}
\label{sign of energy}
\end{equation}
where $E_{c}$ and $\sigma$ are the fit center and the fit error of the energy spectrum (see Fig.~\ref{Experiment sequence}d).
Fig.~\ref{Result}b gives a clear representation of the topological
phase transition by measuring $\nu$ versus $\mu$, where a sharp
change of $\nu$ occurs near $\mu\approx-1.3$. The deviation of the
critical point from the theoretical expectation value $\mu=-1.29$ is
due to the inaccuracy of measuring the very small eigenenergy (close
to 0) near the critical point for which even a slight shift will change
its sign. This deviation can be eliminated by using a NV sample with longer coherence time. Away from the critical point, the measured topological number will only have a negligibly small deviation from the exact value.

In conclusion, we have demonstrated quantum simulation of a topological phase transition with a single NV center at room temperature. Using quantum algorithm of finding eigenvalues, we can not only obtain the dispersion relations, but also directly extract the topological number of the system. Different from the scheme of integration of dynamic responses \cite{Gritsev12,Schroer2014,roushan2014}, our approach of direct measurement of topological number can unambiguously give a discretized value of $\nu$ over almost all parameter space except for a small region around the phase transition. Even in the presence of large magnetical field fluctuations that may smear out the energy gap in the dispersion relations, the approach of direct extraction of topological number remains robust and unambiguously characterizes the topological phase transition.

We may further improve our NV-center-based quantum simulators by using isotopically purified diamond, with significantly extended electron spin coherence time \cite{ultralong2009}. Moreover, with reliable control of multiple spins of the NV center \cite{Bonato15}, more complicated topological systems can be simulated. Utilizing entanglement can lead to a scalable quantum simulator of NV centers \cite{Yao12a}. In addition, the quantum algorithm of finding eigenvalues can be extremely efficient for multiple spins, with only a polynomial time overhead with the number of spins \cite{Abrams99}. Therefore, the NV-center-based quantum simulator is a very promising platform, which will provide a powerful tool to investigate novel quantum system.

\begin{methods}
\subsection{Measurement of $a_{\psi,m}$.}

For the state:
\[
\left ( \left | 6 \right \rangle + \sum_{l=4,5,7,8}a_{l,m} \left | l \right \rangle \right )/\sqrt{2},
\]
if we apply a $\pi/2$ manipulation between $\left | 6 \right \rangle$ and $\left | \Psi \right \rangle$ along -x axis with phase $\theta$, the state will be
\[
\frac{1-a_{\psi,m}e^{i\theta}}{2}\left | 6 \right \rangle + \frac{a_{\psi,m}+e^{-i\theta}}{2}\left | \psi \right \rangle + \sum_{l \neq \psi}a_{l,m} \left | l \right \rangle /\sqrt{2}.
\]
The photoluminescence (PL) of this state is
\begin{equation}
\begin{aligned}
PL(\theta)&=|\frac{1-a_{\psi,m}e^{i\theta}}{2}|^{2}PL_{6}+|\frac{a_{\psi,m}+e^{-i\theta}}{2}|^{2}PL_{\psi}+\sum_{l \neq \psi}|\frac{a_{l,m}}{\sqrt{2}}|^{2}PL_{l}\\
&=\frac{1}{2}(PL_{\psi}-PL_{6})(Re\{a_{\psi,m}\}cos\theta+Im\{a_{\psi,m}\}sin\theta)+C,
\end{aligned}
\end{equation}
where $PL_{l}$ ($l=4,5,6,7,8$) is the photoluminescence of pure states $\ket{l}$, $C$ is a constant independent on $\theta$. With different RF phases $\theta$, a set of photoluminescence $PL(\theta)$ can be detected and fit with function $y(\theta)=y_{0}+A cos(\theta-\theta_{0})$. Then $a_{\psi,m}$ can be obtained with:
\begin{equation}
\begin{aligned}
a_{\psi,m}=\frac{2A cos(\theta_{0})}{PL_{\psi}-PL_{6}}+i\frac{2A sin(\theta_{0})}{PL_{\psi}-PL_{6}}.
\end{aligned}
\end{equation}
In the case of $\left | \Psi \right \rangle = \left | 5 \right \rangle$, an additional $\pi$ pulse between $\left | 6 \right \rangle$ and $\left | 5 \right \rangle$ is applied to make the oscillation amplitude more larger because $PL_4 > PL_5 \approx PL_6$.

\end{methods}


\bibliography{Maintextref}


\begin{addendum}
\item [Acknowledgements]
This work was supported by the National Key Basic Research Program of China (Grant No. 2013CB921800), the National Natural Science Foundation of China (Grant Nos. 11227901, 91021005, 11104262, 31470835), and the Strategic Priority Research Program (B) of the CAS (Grant No. XDB01030400). LJ acknowledges the support from ARL-CDQI, ARO (Grant Nos. W911NF-14-1-0011, W911NF-14-1-0563), AFOSR MURI (Grant Nos. FA9550-14-1-0052, FA9550-14-1-0015), Alfred P. Sloan Foundation (Grant No. BR2013-049), the Packard Foundation (Grant No. 2013-39273).

\item[Author contributions]
J.D. and L.J. proposed the idea. C.J. and Y.L. designed the experimental proposal. F.K., X.K. and P.W. prepared the experimental set-up. M.W. prepared the diamond sample. F.K. and Y.L. performed the experiments. J.D. supervised the setup and experiments. C.J., F.K. and C.L. carried out the calculation. L.J., C.L., C.J., and F.K. wrote the paper. All authors analysed the data, discussed the results and commented on the manuscript.

\item[Competing Interests]
The authors declare that they have no competing financial interests.

\item[Correspondence]
Correspondence and requests for materials
should be addressed to Liang Jiang and Jiangfeng Du (email: liang.jiang@yale.edu; djf@ustc.edu.cn).
\end{addendum}

\newpage

\linespread{1.1}{
\begin{figure}
\centering
\includegraphics[width=0.95\columnwidth]{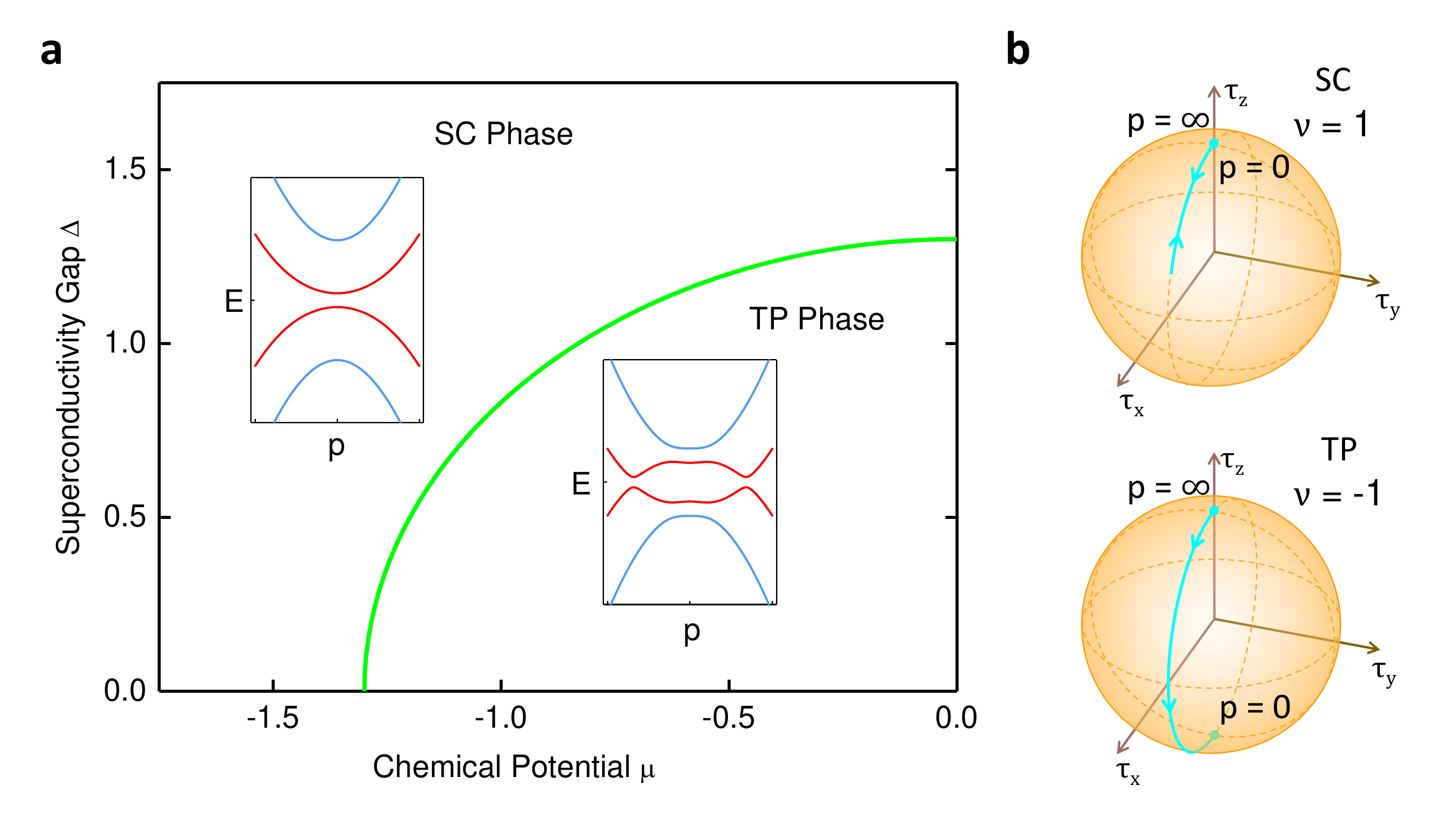}
    \caption{\textbf{Phase diagram and geometric illustration of the topologically distinct phases}.
\textbf{a}, Phase diagram of quantum wire system (calculated at $B_x = 1.3$).
The green line gives the boundary between SC phase and TP phase. The energy dispersion relations of an SC point and a TP point are plotted in the insets.
\textbf{b}, Geometric illustration of topological difference between the SC and the TP phases. }
\label{Phase_diagram}
\end{figure}

\begin{figure}
\centering
\includegraphics[width=0.95\columnwidth]{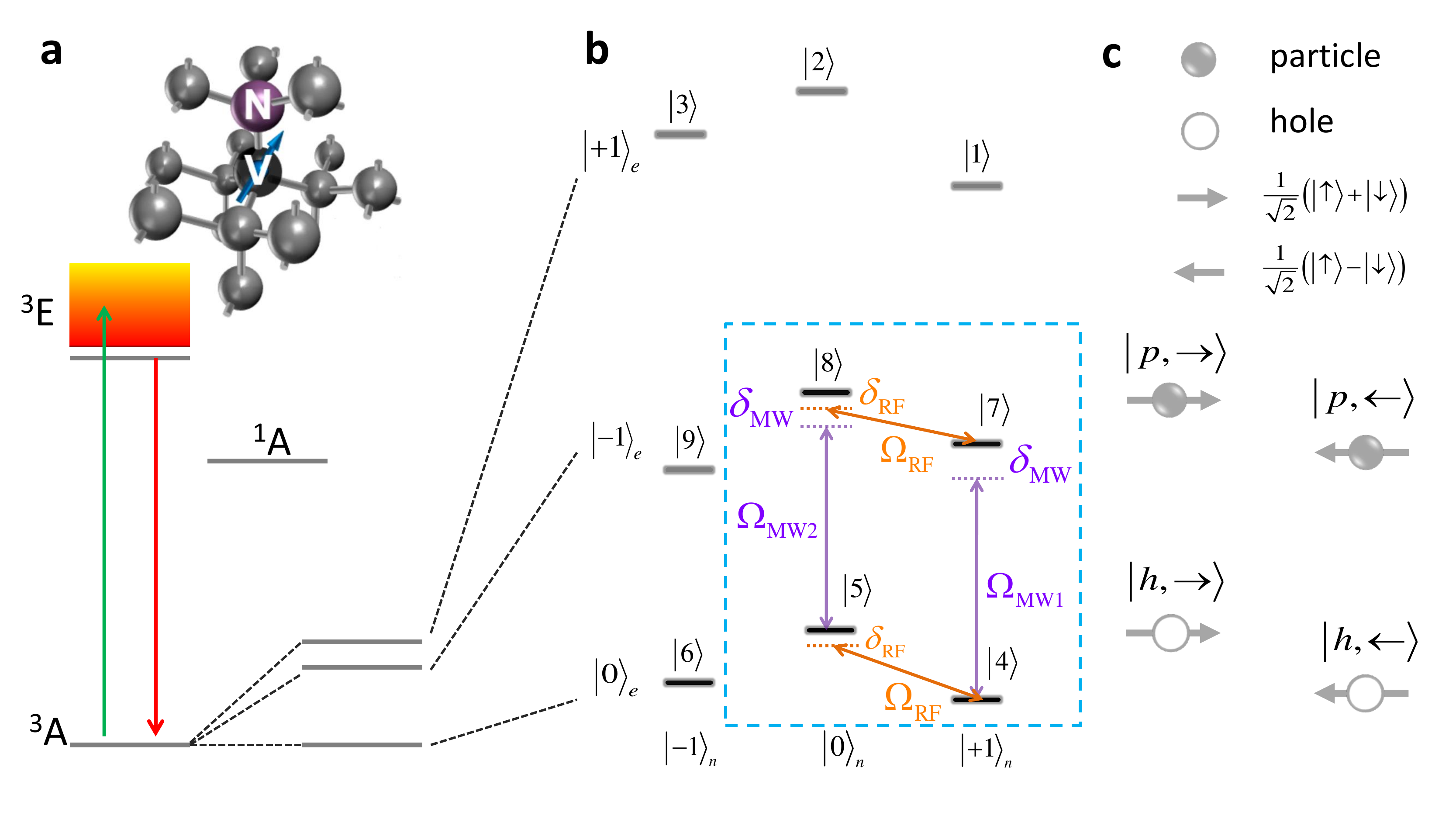}
  \caption{
   \textbf{NV system and its correlation with QW system}.
   \textbf{a}, Structure and energy levels of the NV centers.
    \textbf{b}, Hyperfine structure of the coupling system with NV electron spin and $^{14}$N nuclear spin. The 9 energy levels are labeled as $\left | 1 \right \rangle$ to $\left | 9 \right \rangle$. The quantum simulation is carried out in the subspace spanned by $\{ \left | 4 \right \rangle,\left | 5 \right \rangle,\left | 7 \right \rangle,\left | 8 \right \rangle \}$. Two MW pulses (purple arrows) and two RF pulses (orange arrows) are applied simultaneously to selectively drive the corresponding electron and nuclear spin transitions.
    \textbf{c}, Four basis states of QW system corresponding to the four NV states inside the square box.}
  \label{Protocol}
\end{figure}

\begin{figure}
\centering
\includegraphics[width=0.95\columnwidth]{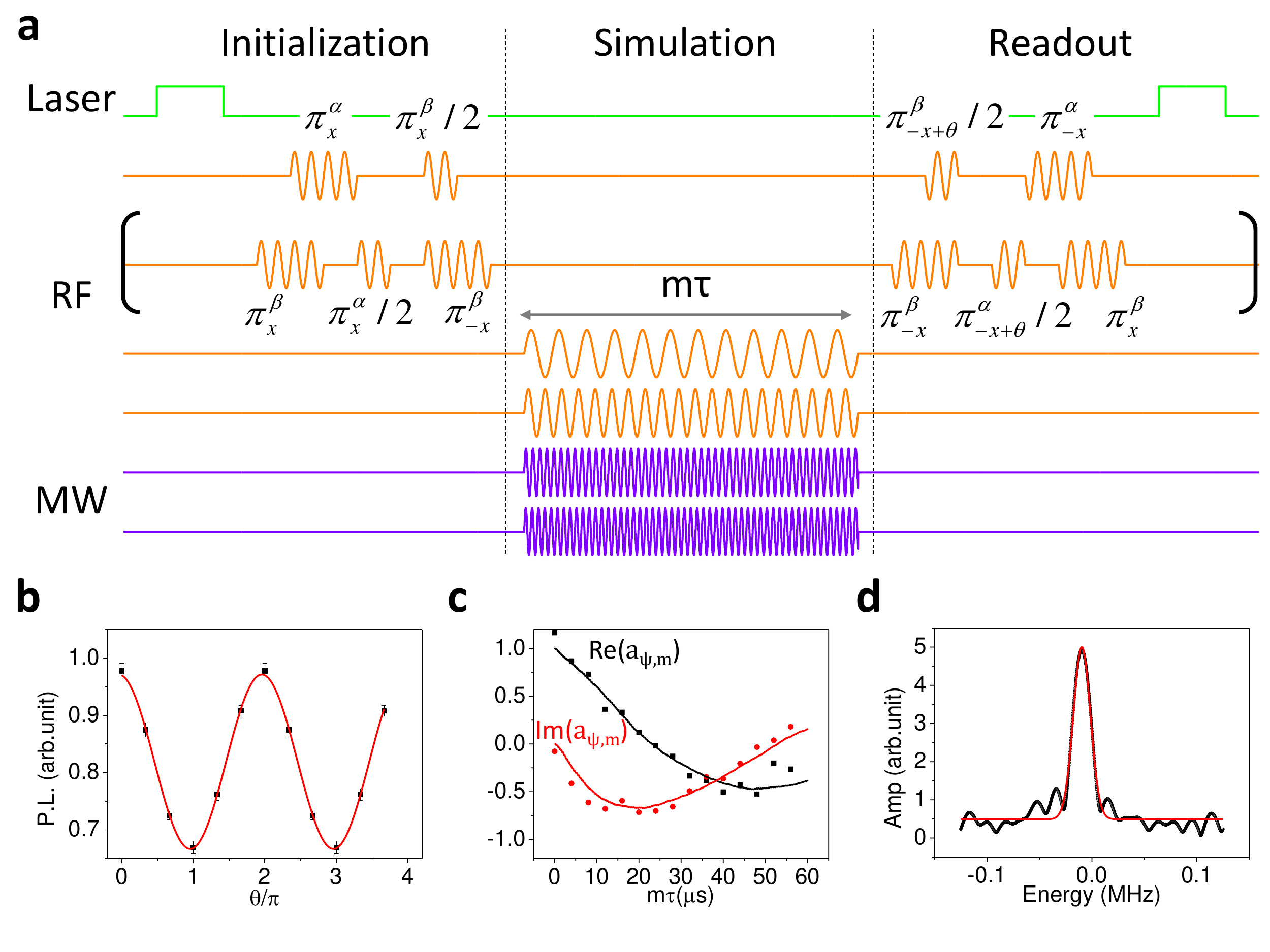}
    \caption{\textbf{Simulation QW Hamiltonian and detection its eigenvalues.}
\textbf{a}, The pulse scheme. The superscript $\alpha$ ($\beta$) of the RF pulses indicates the nuclear spin operation between $\left | 4 \right \rangle$ and $\left | 5 \right \rangle$ ($\left | 5 \right \rangle$ and $\left | 6 \right \rangle$). For the initialization and readout parts there are two pulse sequences, which corresponds to the two initial state cases $\left | \Psi \right \rangle = \left | 5 \right \rangle$ (the upper pulse sequence) and $\left | 4 \right \rangle$ (the lower bracketed pulse sequence), respectively.
\textbf{b}, Photoluminescence changes versus different RF $\pi/2$ pulse phase $\theta$ for fixed evolution time $m\tau$. The points are the experimental data and the curve is the sine function fit. Error bars indicate $\pm1$ s.d. induced by the photon shot noise.
\textbf{c}, Measurement of $a_{\psi,m}$ with different evolution time $m\tau$. Black and red lines are the numerical calculation results.
\textbf{d}, The energy spectrum of the simulated Hamiltonian yielded from the Fourier transform of the time-domain data in \textbf{c}. A Gauss function fit (the curve) is performed to get the exact eigenenergy value. }
\label{Experiment sequence}
\end{figure}

\begin{figure}
\centering
\includegraphics[width=0.95\columnwidth]{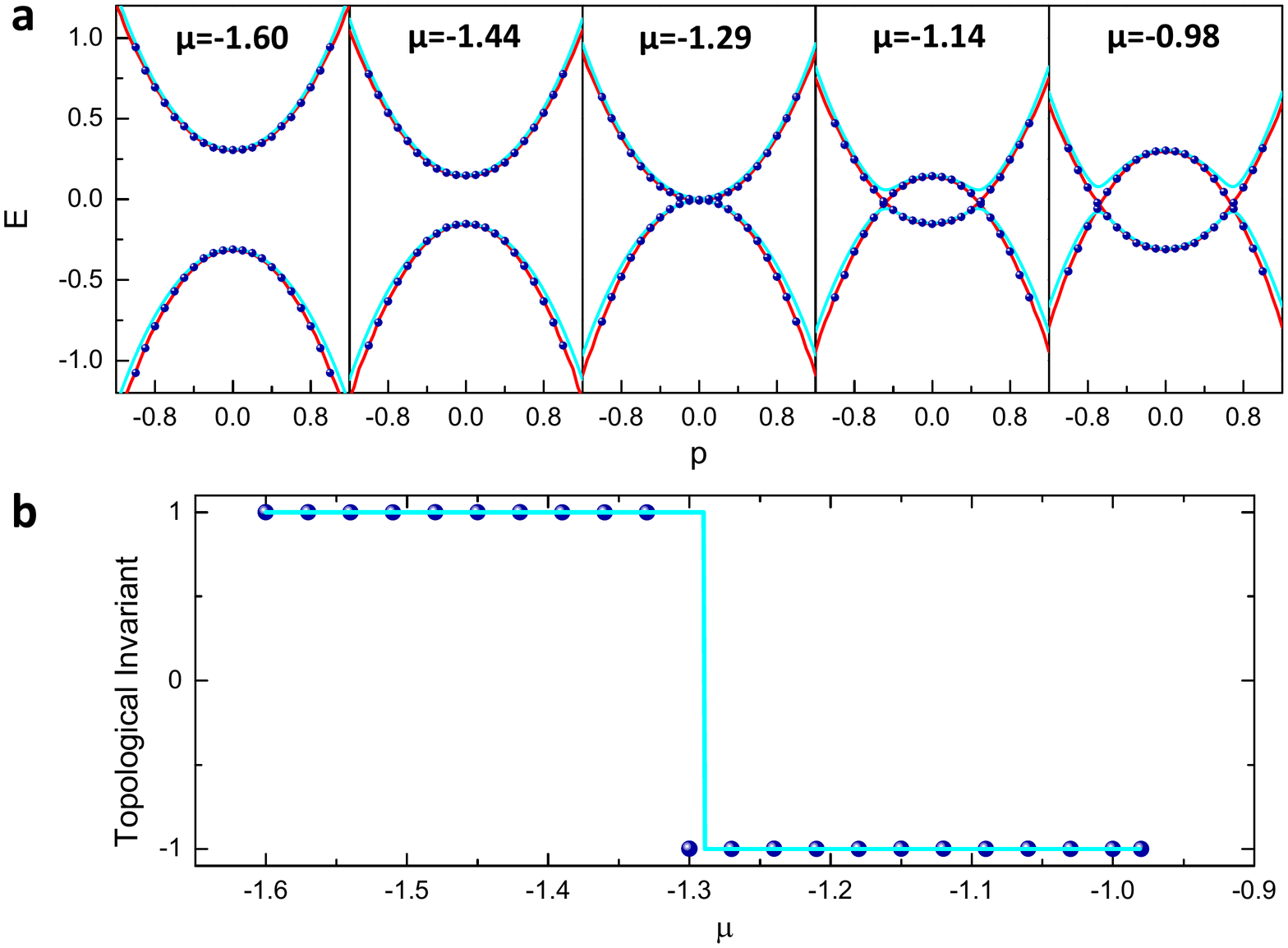}
    \caption{\textbf{Energy dispersion relations and topological phase transition.}
\textbf{a}, Energy dispersion relations with different chemical potential $\mu$. The points, light cyan lines, and red lines represent the experimental, analytical, and numerical results, respectively. Error bars given by fit error are smaller than the symbols. As the energy bonds are symmetrical about $p=0$, only the right half points (i.e. $p\geqslant 0$) are actually measured in the experiment.
\textbf{b}, The measured topological number $\nu$ versus the chemical potential $\mu$, which shows a topological phase transition happened near $\mu \approx -1.3$. The cyan line is the theoretical prediction. }
\label{Result}
\end{figure}
}

\end{document}